\documentclass[12pt,a4paper]{article}
\usepackage[affil-it]{authblk}
\usepackage{graphicx} % Required for inserting images.
\usepackage[margin=25mm]{geometry}
\usepackage[font=sf]{caption} % Changes font of captions.

\usepackage{tikz}
\usetikzlibrary{positioning, calc}
\usepackage{pgfplots}
\usepackage{ragged2e}
\pgfplotsset{compat=1.17}
\usepackage{amsmath}
\usepackage{amssymb}
\usepackage{booktabs}
\usepackage{siunitx}
\usepackage{url}

\definecolor{mygreen}{RGB}{220,240,235} % A shade of green
\definecolor{mylightblue}{RGB}{95,147,160}  % A shade of blue
\definecolor{myorange}{RGB}{255,152,0} % A shade of orange

\title{Unsupervised Physics-Informed Deep Learning for Dual-Energy CT Material Decomposition}

\author[1,2]{Laura Hellwege}
\author[2]{Johann Christopher Engster}
\author[2]{Moritz Schaar}
\author[1,2]{Thorsten M. Buzug}
\author[2]{Maik Stille}
\affil[1]{\small Institute of Medical Engineering, University of Lübeck, Lübeck, Germany}
\affil[2]{\small Fraunhofer Research Institution for Individualized Medical Technology and Engineering (IMTE), Lübeck, Germany}

\begin{document}
\maketitle

\begin{abstract}
\justifying
Dual-energy computed tomography (DECT) enables material-specific imaging through acquisitions at two different X-ray energy spectra. Material decomposition from DECT data is an ill-posed inverse problem that is highly sensitive to noise amplification. 
Conventional methods face challenges regarding accuracy and computational efficiency. 
We present a novel physics-informed deep learning (DL) framework for DECT material decomposition that eliminates the requirement for ground-truth material images during training. Our approach incorporates a polychromatic forward model into the training pipeline, enabling the network to learn the decomposition mapping by minimizing discrepancies in the projection domain. We validate our method on the AAPM DL-Spectral CT Challenge dataset, comparing performance against three state-of-the-art methods. In the projection domain, our method achieves the lowest root mean squared error (RMSE) across test datasets. 
For virtual monoenergetic images (VMIs) at 30\,keV, 50\,keV, and 70\,keV, the approach consistently outperforms all conventional methods in both RMSE and structural similarity index (SSIM). 
These results demonstrate the potential of DL for accurate material decomposition in DECT without requiring labeled training data.
\end{abstract}

\section{Introduction}

Dual-energy computed tomography has become an established clinical imaging modality. It exploits the energy-dependent attenuation properties of materials to enable quantitative tissue characterization beyond conventional single-energy CT \cite{mccollough2015dual}. By acquiring projection data at two different X-ray energy spectra, low energy (LE) and high energy (HE) spectra, DECT provides sufficient information for decomposing the measured attenuation into contributions from basis materials, typically water and iodine or bone and soft tissue \cite{alvarez1976energy}. 
This capability enables various clinical applications including virtual non-contrast imaging, iodine quantification, bone removal, and the generation of virtual monoenergetic images (VMIs) for artifact reduction and optimized contrast \cite{johnson2007material, yu2012virtual}.

The fundamental principle underlying DECT material decomposition is that the linear attenuation coefficient of any material can be approximated as a linear combination of the attenuation coefficients of two basis materials across the diagnostic energy range. Given measurements at two distinct energy spectra, this relationship enables solving for the concentrations or area densities of the basis materials at each spatial location. However, material decomposition constitutes an ill-posed inverse problem that amplifies noise present in the acquired projections  \cite{mccollough2015dual}. 
%This noise amplification is particularly pronounced when decomposing into materials with similar attenuation characteristics or when operating at low radiation doses.

Several approaches have been developed to address the material decomposition problem. Image-domain methods perform decomposition on reconstructed CT images, utilizing empirical calibration relationships between CT image values at different energies and material concentrations \cite{stenner2007empirical, liu2009quantitative}. While computationally efficient, these methods are affected by beam-hardening artifacts present in the reconstructed images and may not fully exploit spectral information. 
Projection-domain methods decompose the raw projection data before reconstruction, potentially avoiding beam-hardening artifacts but requiring accurate knowledge of the X-ray spectra and detector response \cite{stenner2007empirical, hellwege2023aedec}. 
%Hybrid approaches combine aspects of both domains \cite{niu2014iterative}.
Iterative methods formulate material decomposition as an optimization problem that simultaneously enforces data consistency and applies regularization to mitigate noise amplification \cite{long2014multi, mechlem2017joint}. These methods can achieve high accuracy but are computationally expensive, requiring iterative forward and backward projections through a polychromatic model.
Deep learning (DL) has emerged as a powerful approach for medical image reconstruction and analysis. For DECT material decomposition, supervised DL methods have demonstrated promising results by learning the mapping from dual-energy data to material images using paired training data \cite{clark2018deep, xu2021image}. 

However, supervised training is fundamentally limited by the fact that ground-truth material images cannot be directly measured in clinical practice, since material fractions are abstract decomposition coefficients with no physical measurement equivalent. Only derived quantities such as virtual monoenergetic images can be validated against physical reference data.
% However, obtaining ground-truth material images for supervised training is challenging in clinical practice. 
%, as they require either phantom measurements, simulations, or alternative gold-standard acquisitions.
Recently, unsupervised and unsupervised DL approaches have gained attention for their ability to train networks without explicit ground-truth labels \cite{chen2022self, hendriksen2020noise2inverse}. In our previous work, we demonstrated an unsupervised DL framework for CT reconstruction that leverages geometry-informed training through projection operators \cite{hellwege2024unsupervised}. The network learns to reconstruct images by minimizing discrepancies in the projection domain, eliminating the need for ground-truth images during training.\\

In this work, we extend and adapt the unsupervised framework to the more complex problem of DECT material decomposition. The key challenge lies in incorporating the polychromatic forward model, which relates material images to the measured dual-energy projections. We validate our approach on the AAPM DL-Spectral CT Challenge dataset \cite{aapm_spectral}, comparing against established methods including empirical dual-energy calibration (EDEC), its advanced variant (AEDEC), and model-based iterative reconstruction (MBIR). This is achieved without using a large dataset of ground truth images. Instead, our method requires only the measured dual-energy projection data, knowledge of the X-ray spectra, detector response and projection geometry as well as the polychromatic forward model.

\section{Methods}

\subsection{Problem Formulation}

Let $\mu^X(E_k)$ denote the linear attenuation coefficient of material $X$ at energy $E_k$. Let a voxel of the material image $f^X$ be denoted by $f_j^X$, $j \in \{1,\ldots,N\}$. Let the variable $a_{ij}$ denote the entry of the projection matrix $\mathbf{A}$ for voxel $j$ in projection ray $i$. Then the discrete polychromatic model $\mathcal{P}^S$ 
describes the polychromatic projection measurement $p_i^S$ for spectrum $S \in \{\text{LE}, \text{HE}\}$ by
\begin{align}
\label{eq:polychromatic} 
p^S_i &= -\ln\bigg(\mathcal{P}^S(\mathbf{A}f)\big\vert_{i} \bigg) \\
& = -\ln\bigg(\sum_{E_k} S(E_k) \exp\biggl(-\sum_{X} \mu^X(E_k) \sum_{j=1}^N a_{ij}f_j^X \biggr)\bigg) ,
\end{align}
where $S(E_k)$ is an entry of the the normalized source spectrum including detector response.
It is assumed that the attenuation coefficient can be decomposed into contributions from two basis materials $A$ and $B$ with linear attenuation coefficients $\mu^A(E)$ and $\mu^B(E)$:
\begin{equation}
\mu_j^X(E_k) = f_j^A \mu^A(E_k) + f_j^B \mu^B(E_k),
\label{eq:decomposition}
\end{equation}
where $f^A_j$ and $f^B_j$ are the linear attenuation contributions of the two basis materials. The goal of material decomposition is to recover the spatial distributions $f^A$ and $f^B$ from the measured projections $p^{\text{LE}}$ and $p^{\text{HE}}$.

\subsection{Conventional Methods}

The Empirical Dual-Energy Calibration (EDEC) method establishes an empirical mapping between measured projection values and material images through calibration measurements on phantoms with known material compositions \cite{stenner2007empirical}. The operator $\mathcal{O}^\mathcal{C}_{\text{EDEC}}$, whose coefficients $\mathcal{C}$ are determined by calibration, directly estimates material images $\hat{f}^A, \hat{f}^B$ from the dual-energy projections $p^{\text{LE}}, p^{\text{HE}}$, i.e.
\begin{equation}
(\hat{f}^A, \hat{f}^B) = \mathcal{O}^\mathcal{C}_{\text{EDEC}}(p^{\text{LE}}, p^{\text{HE}}).
\end{equation}
While computationally efficient once calibrated, EDEC accuracy depends on the calibration phantom adequately representing the range of material combinations encountered in patient imaging.
Advanced EDEC (AEDEC) \cite{hellwege2023aedec} extends the basic EDEC approach by incorporating additional correction terms or refined calibration procedures to improve accuracy, particularly for material combinations that deviate from the calibration range. 
%This may include polynomial corrections, scatter corrections, or adaptive calibration strategies.
Model-Based Iterative Reconstruction (MBIR) formulates material decomposition as an optimization problem:
\begin{equation}
(f^{A\star}, f^{B\star}) = \arg\min_{f^A, f^B} \| p^\text{LE} - \mathcal{P}^{\text{LE}}(f^A, f^B) \|_2^2 + \| p^\text{HE} - \mathcal{P}^{\text{HE}}(f^A, f^B) \|_2^2 + \lambda R(f^A, f^B),
\label{eq:mbir}
\end{equation}
where $\mathcal{P}^S$ represents the polychromatic forward model from equation~\eqref{eq:polychromatic}, and $R$ is a regularization functional with weight $\lambda$. In this work, we employ MBIR without explicit regularization ($\lambda = 0$) to provide a fair comparison with our unregularized DL approach.

\subsection{Unsupervised Deep Learning Framework}

\begin{figure}[t]
    \centering
    \sffamily
    \resizebox{0.85\textwidth}{!}{%
    \begin{tikzpicture}[
        % boxorange/.style={draw=none, fill=orange!20!, text width=1.7cm, text centered, minimum height=2cm, font=\huge},
        % boxblue/.style={draw=none, fill=mylightblue, text width=1.7cm, text centered, minimum height=2cm, font=\huge},
        boxorange/.style={draw=none, fill=white, text width=1.7cm, text centered, minimum height=2cm, font=\huge},
        boxblue/.style={draw=none, fill=white, text width=1.7cm, text centered, minimum height=2cm, font=\huge},
        boxgrey/.style={draw=gray!30, fill=gray!10!white, text width=1cm, text centered, minimum height=4cm, font=\huge},
        ellipseboxorange/.style={draw=orange, circle, minimum height=1cm, minimum width=2cm, fill=orange!10, font=\huge},   ellipseboxgray/.style={draw=gray!30, circle, minimum height=1cm, minimum width=2cm, fill=gray!10, font=\huge},
        rect/.style={draw, rectangle, minimum height=1cm, minimum width=2cm, fill=orange!20, text centered, font=\huge},
        arroworange/.style={->, line width=1mm, orange, >=stealth},
        arrowgray/.style={->, line width=1mm, mylightblue!75!black, >=stealth},
        arrowgreen/.style={->, line width=1mm, mygreen!75!black, >=stealth},
        arrowgray/.style={->, line width=1mm, gray, >=stealth},
        dashedarroworange/.style={->, line width=1mm, orange, dashed, >=stealth},
        dashedarrowgray/.style={->, line width=1mm, gray, dashed, >=stealth}, 
        node distance=1cm
    ]
    
    % % Left side measurements
    % \node[boxorange] (y) {\huge $ p$};
    % \node[draw=none, above=of y, yshift=0cm, font=\large] (labelMeasurement){Measurements};
    \node[anchor=center, boxorange, minimum width=2.8cm, minimum height=2.8cm] (measurements) {};
    \node[anchor=center] at (measurements) {\includegraphics[trim=15cm 3cm 22cm 3cm, clip, height=2.5cm, width=2.5cm] {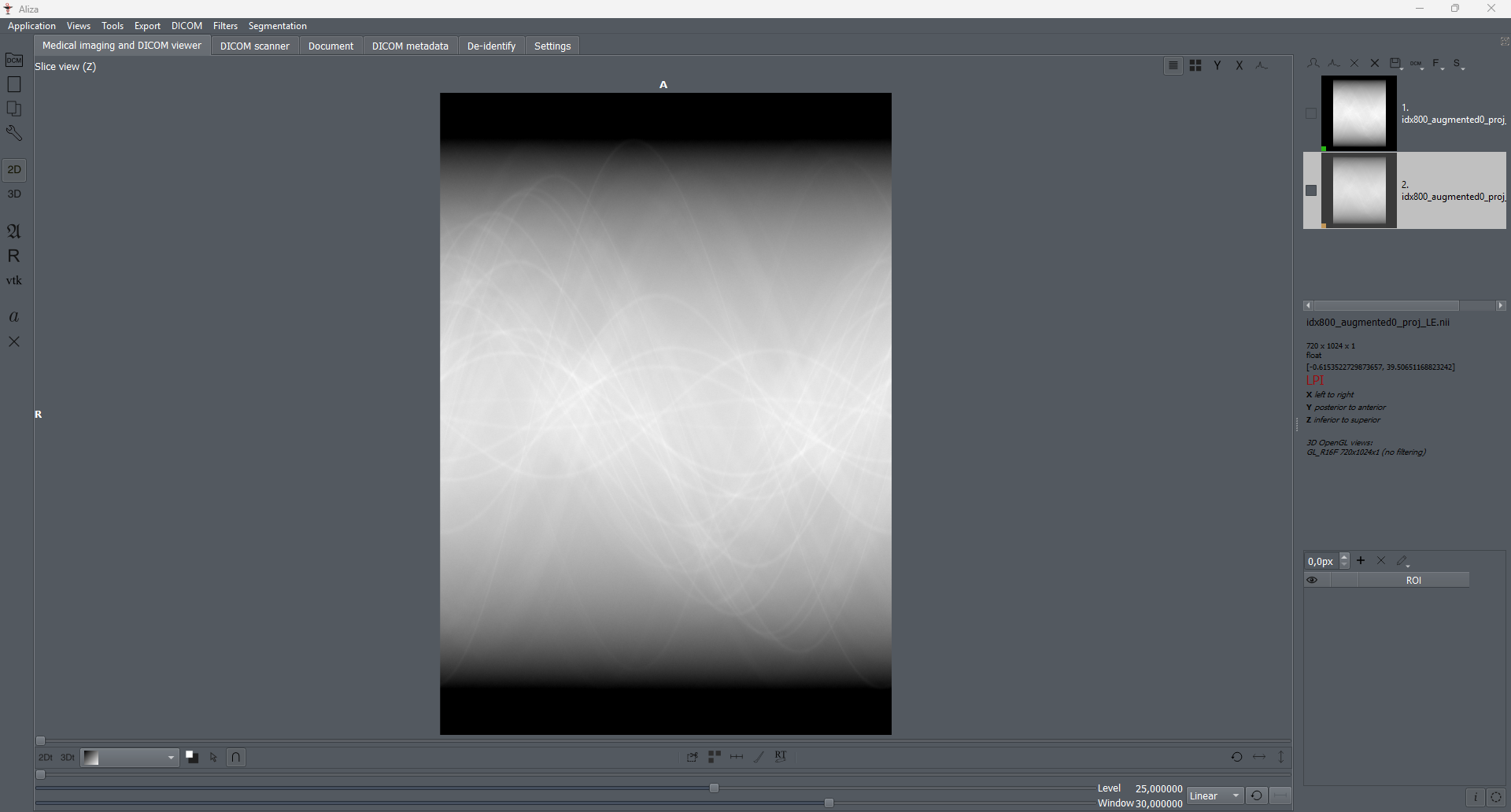}};    
    \node[boxorange, below of=measurements, yshift=-2cm, minimum width=2.8cm, minimum height=2.8cm] (measurements_HE) {};
    \node[anchor=center] at (measurements_HE) {\includegraphics[trim=15cm 3cm 22cm 3cm, clip, height=2.5cm, width=2.5cm] {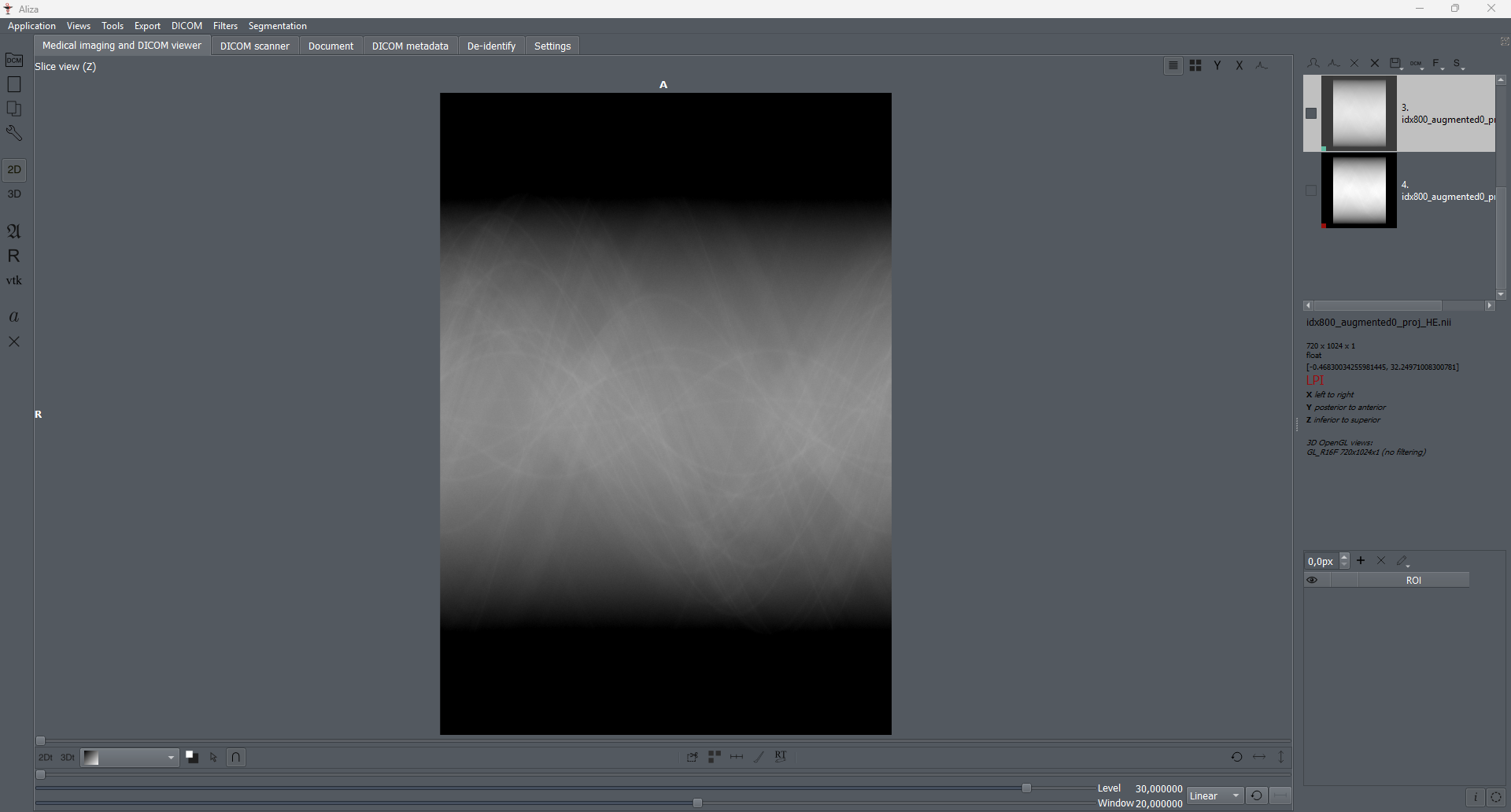}};
    \node[draw=none, above=of measurements, yshift=-0.5cm, font=\large, align=center] (labelMeasurement){Measurement $p$};
    
    % A^T in box
    \node[boxgrey, text width=1.1cm, minimum height=5cm, right=of measurements, xshift=-0.65cm, yshift=-1.5cm] (boxBP) {};
    \node[draw=none, anchor=center] at (boxBP) {\textcolor{gray!75!black}{\Large BP}};

    % Left side BP
    \node[boxblue, right=of measurements, xshift=2.5cm, minimum width=2.8cm, minimum height=2.8cm, anchor=center] (imageBP) {};
    \node[anchor=center] at (imageBP) {\includegraphics[trim=12cm 3cm 18cm 3cm, clip, height=2.5cm, width=2.5cm] {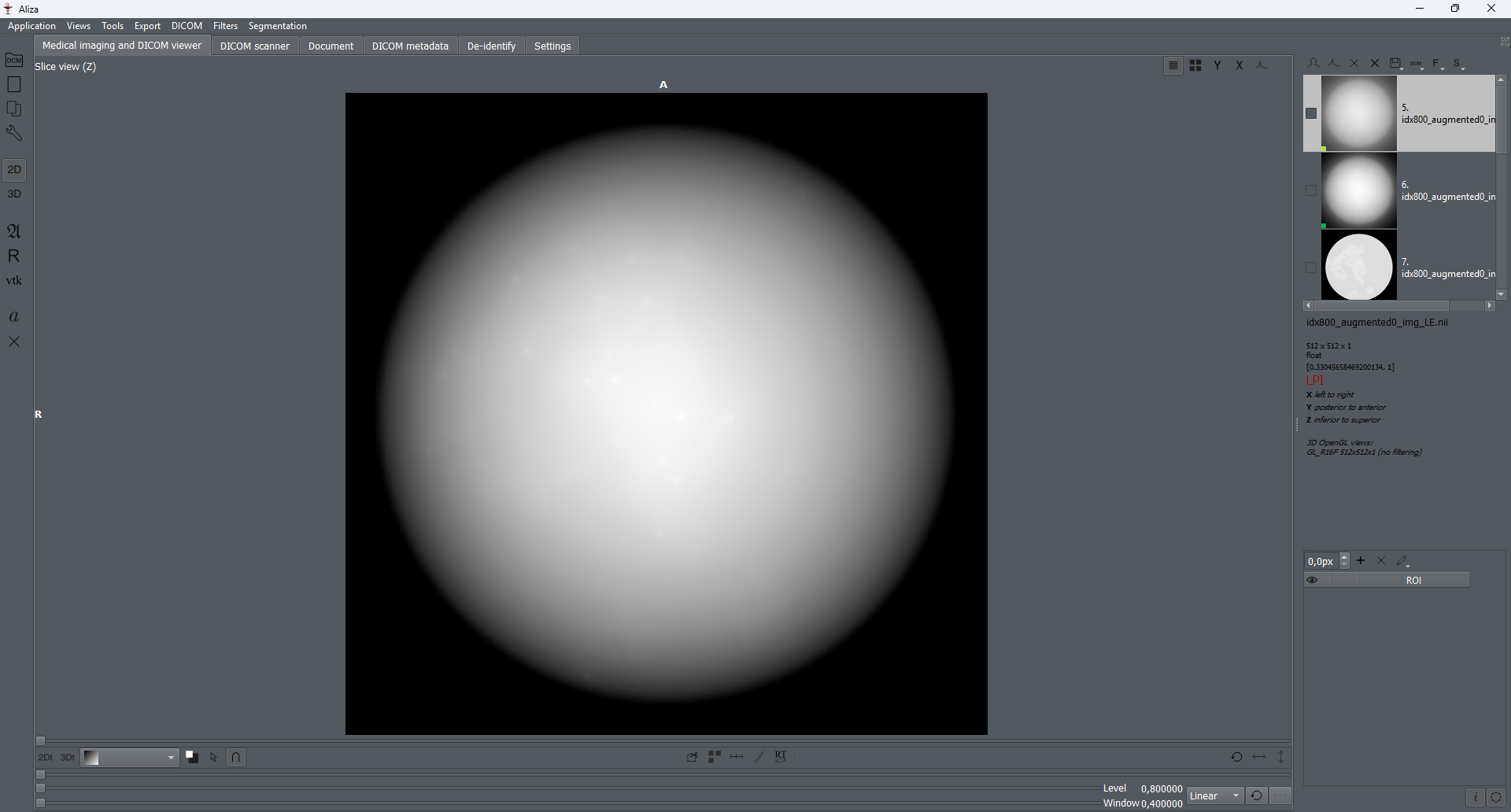}};
    \node[boxblue, below=of imageBP, yshift=-0.6cm, minimum width=2.8cm, minimum height=2.8cm, anchor=center] (imageBP_HE) {};
    \node[anchor=center] at (imageBP_HE) {\includegraphics[trim=12cm 3cm 18cm 3cm, clip, height=2.5cm, width=2.5cm] {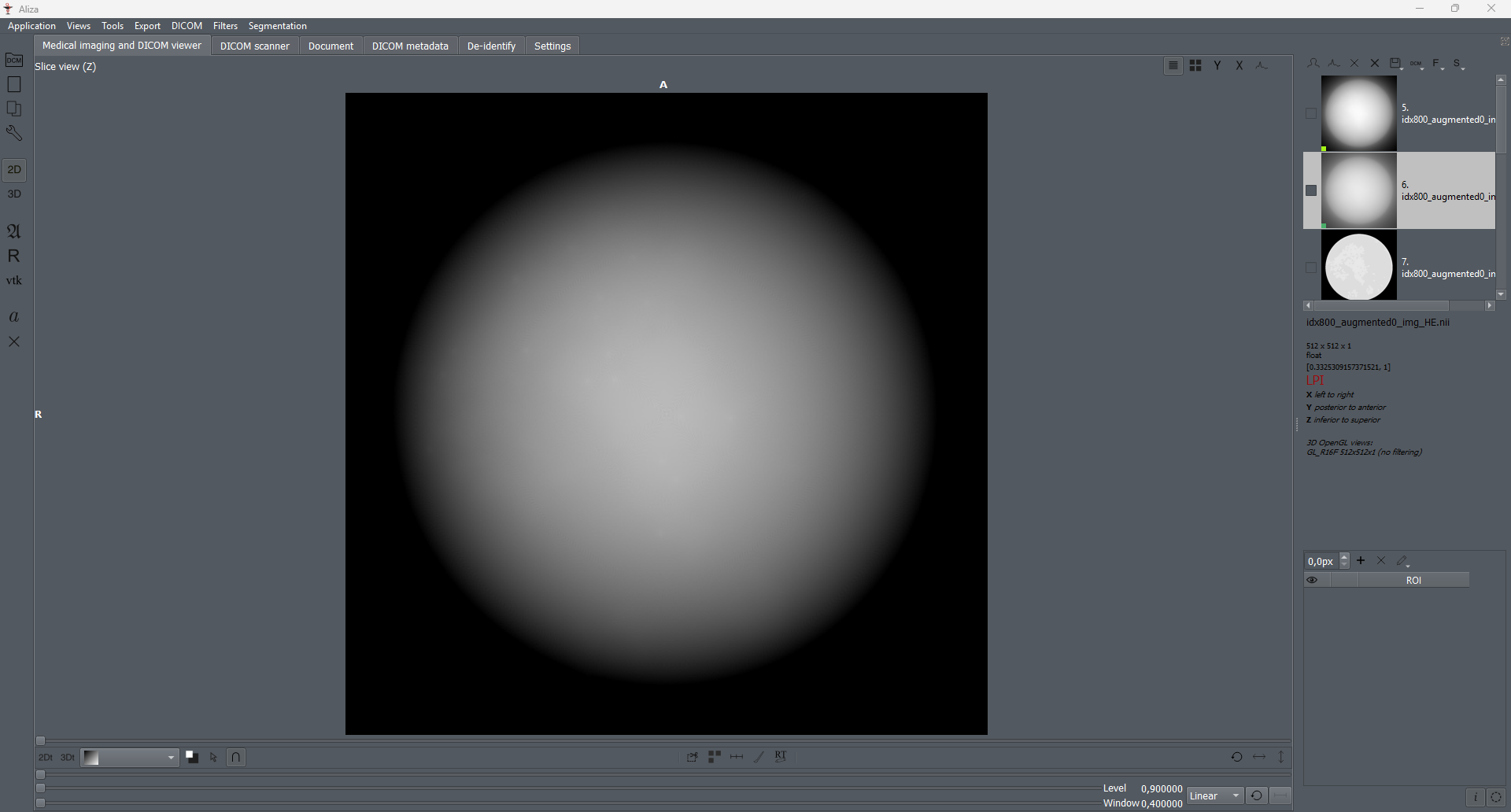}};
    \node[draw=none, above=of imageBP, yshift=-0.5cm, font=\large, align=center] (labelBP) {Simple \\ backprojection $\tilde{f}$};
    \draw[arrowgray] (measurements) -- (imageBP);
    \draw[arrowgray] (measurements_HE) -- (imageBP_HE);

    % Center U-net++
    \node[draw=black, circle, right=of imageBP, xshift=-0.3cm, yshift=-1.5cm, font=\huge] (Unet) {\includegraphics[width=1.2cm]{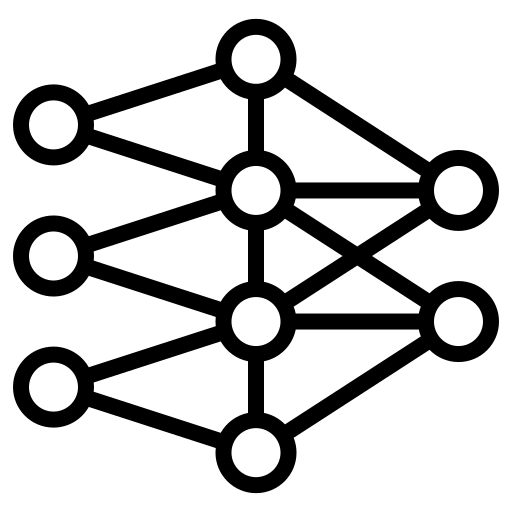}};
    \node[draw=none, minimum size=0.1cm, above=of Unet, yshift=-0.7cm, font=\large] (labelUnet) {UNet++};
    \draw[arrowgray] ($(Unet.180)-(0.75cm,0)$) -- (Unet);
    
    % Right side Images 
    \node[boxblue, right=of imageBP, xshift=4cm,  minimum width=2.8cm, minimum height=2.8cm, anchor=center] (image) {};
    \node[anchor=center] at (image) {\includegraphics[trim=12cm 3cm 18cm 3cm, clip, height=2.5cm, width=2.5cm] {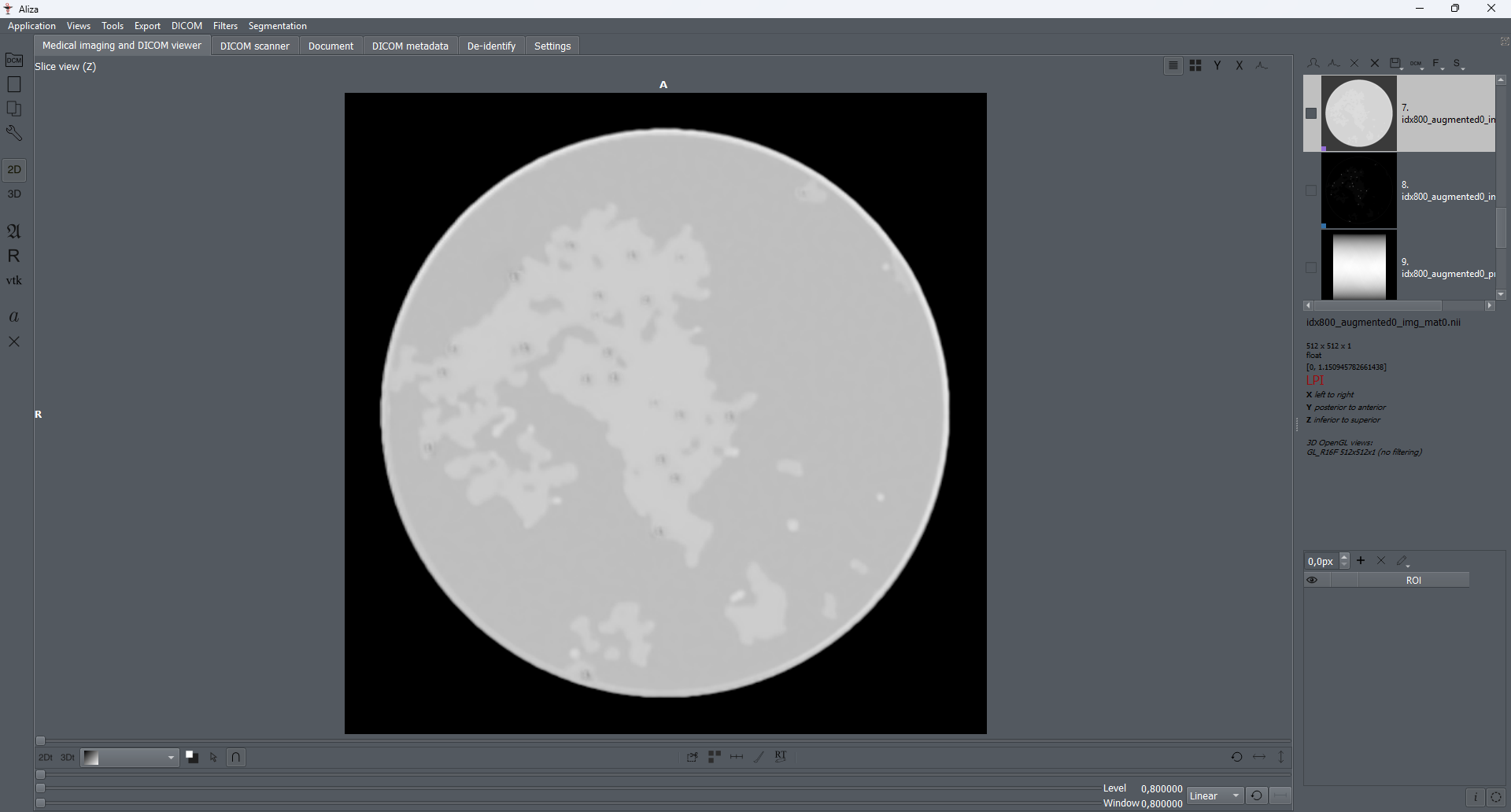}};
    \node[boxblue, below=of image, yshift=-0.6cm, minimum width=2.8cm, minimum height=2.8cm, anchor=center] (image_mat2) {};
    \node[anchor=center] at (image_mat2) {\includegraphics[trim=12cm 3cm 18cm 3cm, clip, height=2.5cm, width=2.5cm] {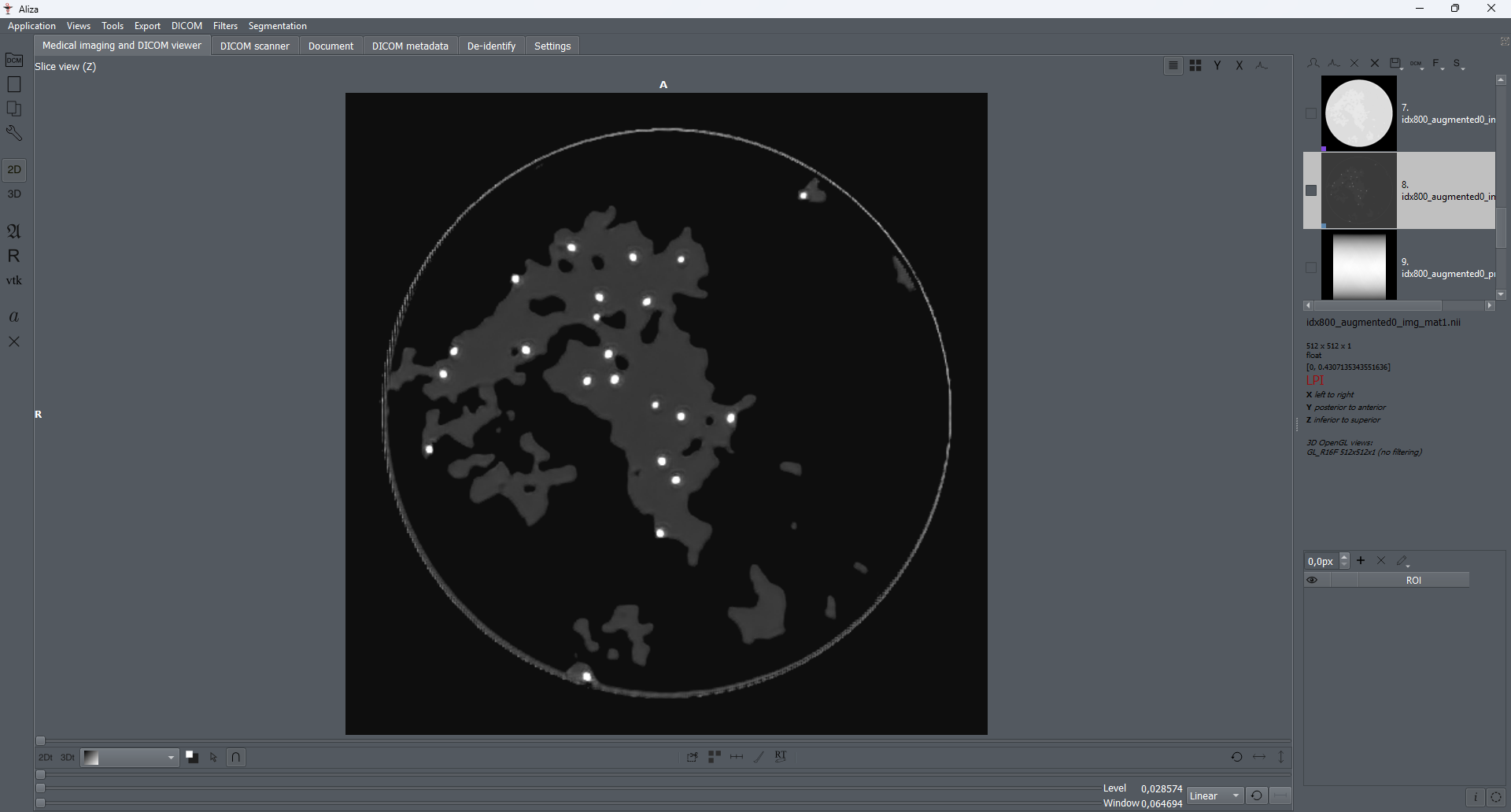}};
    \node[draw=none, above=of image, font=\large, align=center, anchor=center] (labelPred) {Estimation of \\ material images $\hat{f}$};
    \draw[arrowgray] (Unet.10) -- ($(Unet.10) + (0.75,0)$);
    \draw[dashedarroworange] ($(Unet.-10) + (0.75,0)$) -- (Unet.-10);
    
    % Projection
    % \node[boxorange, right=of image, xshift=1.5cm] (proj) {\huge $ \hat{p}$};
    % \node[draw=none, above=of proj, yshift=-0.4cm, font=\large, align=center] {Projections \\ of prediction};
    \node[boxorange, right=of image, minimum width=2.8cm, minimum height=2.8cm, xshift=2.5cm, anchor=center] (proj) {};
    \node[anchor=center] at (proj){\includegraphics[trim=15cm 3cm 22cm 3cm, clip, height=2.5cm, width=2.5cm] {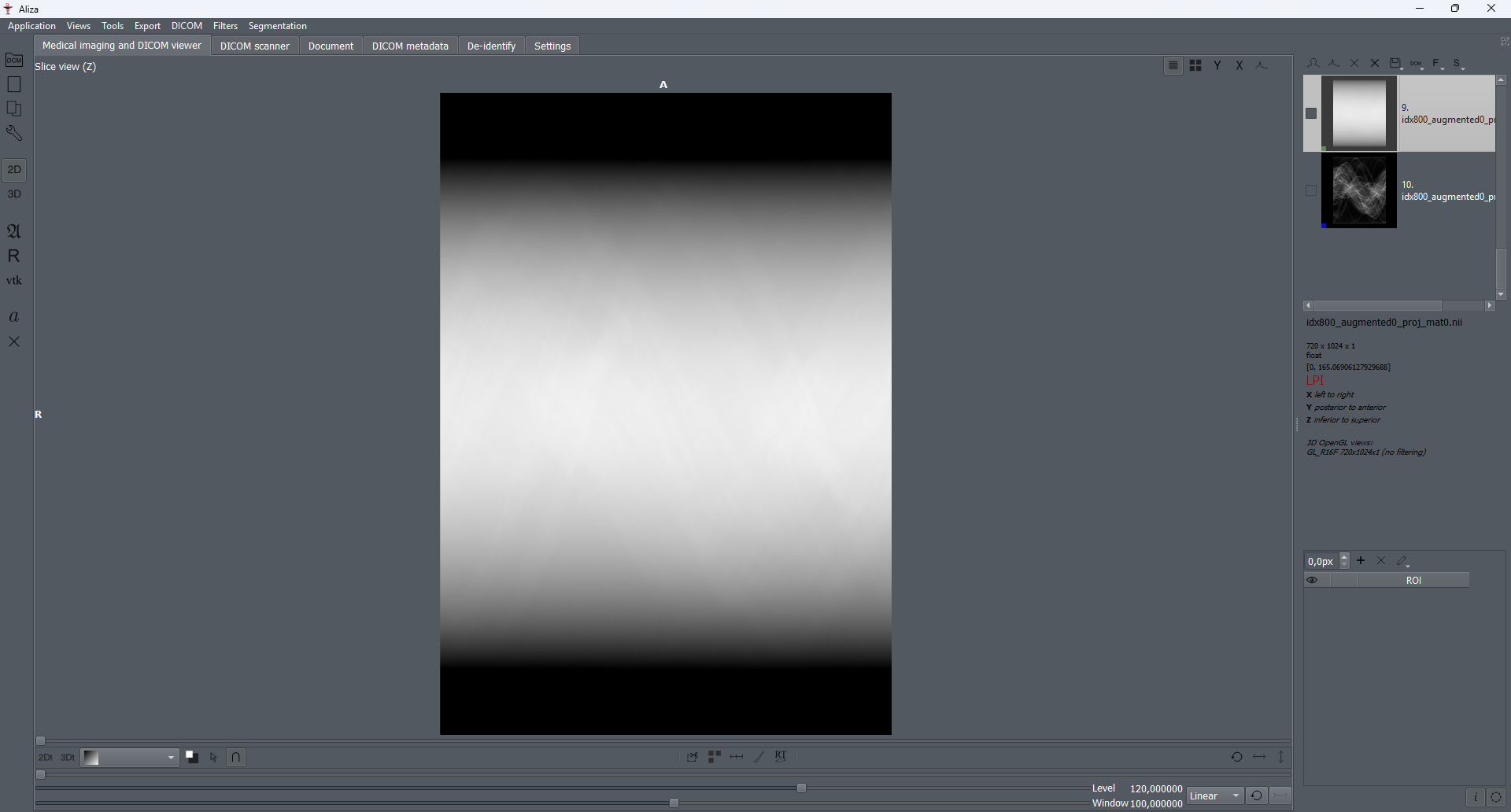}};
    \node[boxorange, below=of proj, yshift=-0.6cm, minimum width=2.8cm, minimum height=2.8cm, anchor=center] (proj_mat2) {};
    \node[anchor=center] at (proj_mat2){\includegraphics[trim=15cm 3cm 22cm 3cm, clip, height=2.5cm, width=2.5cm] {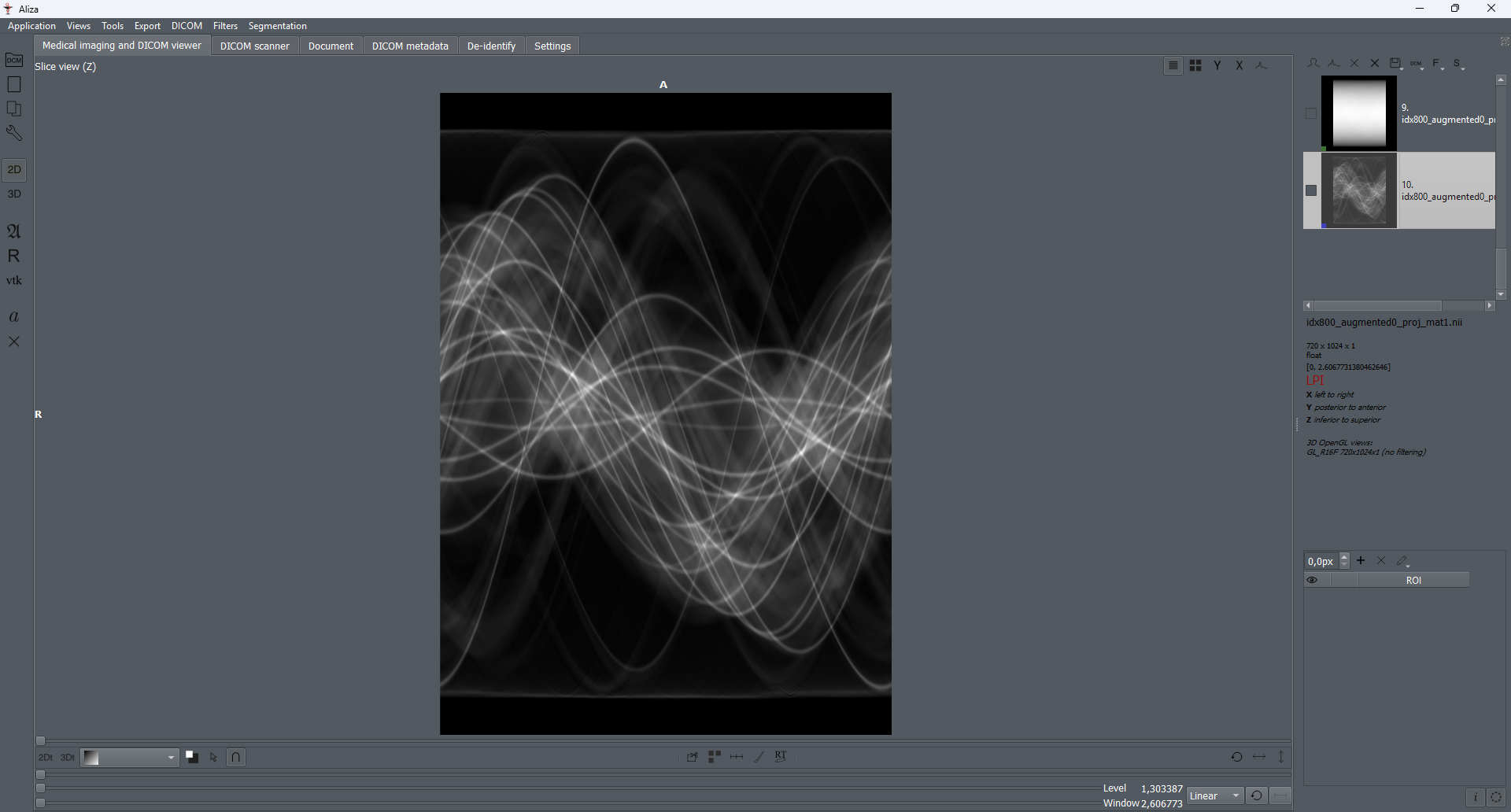}};
    \node[draw=none, above=of proj, font=\large, align=center, anchor=center] {Forward projection \\ of material images};

    % A in box
    \node[boxgrey, text width=1.1cm, minimum height=5cm, right=of image, xshift=-0.65cm, yshift=-1.5cm](Abox) {};
    \node[draw=none, yshift=+0.4cm] at (Abox) {\textcolor{gray!75!black}{\Large FP}};
    \draw[arrowgray] (image.10) -- (proj.170);
    \draw[arrowgray] (image_mat2.10) -- (proj_mat2.170);
    \draw[dashedarroworange] (proj.190) -- (image.-10);
    \draw[dashedarroworange] (proj_mat2.190) -- (image_mat2.-10);
    \node[draw=none, yshift=-0.4cm] at (Abox) {\textcolor{orange!90!black}{\Large BP}};

    % Polychromatic model
    \node[ellipseboxgray, right=of Unet, xshift=7.75cm, yshift=-3.0cm] (Model) {\Large $\mathcal{P}$};
    \draw[arrowgray] (proj.-42) -| (Model.80);
    \draw[dashedarroworange] (Model.100)  |-  (proj_mat2.42);

    % Estimation
    \node[boxorange, below=of proj_mat2, yshift=-1cm, minimum width=2.8cm, minimum height=2.8cm, anchor=center] (proj_LE) {};
    \node[anchor=center] at (proj_LE){\includegraphics[trim=15cm 3cm 22cm 3cm, clip, height=2.5cm, width=2.5cm] {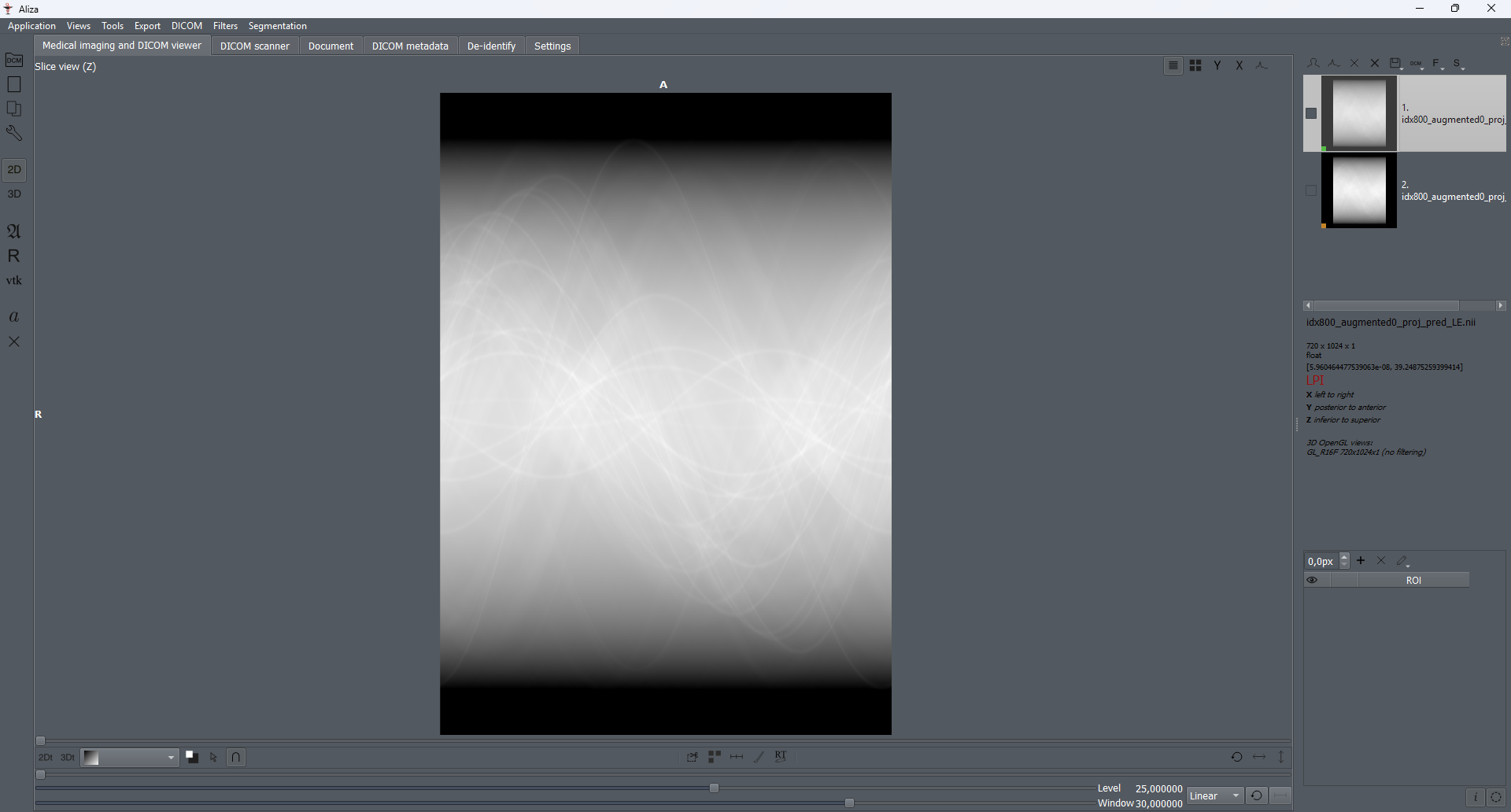}};
    \node[boxorange, below=of proj_LE, yshift=-0.6cm, minimum width=2.8cm, minimum height=2.8cm, anchor=center] (proj_HE) {};
    \node[anchor=center] at (proj_HE){\includegraphics[trim=15cm 3cm 22cm 3cm, clip, height=2.5cm, width=2.5cm] {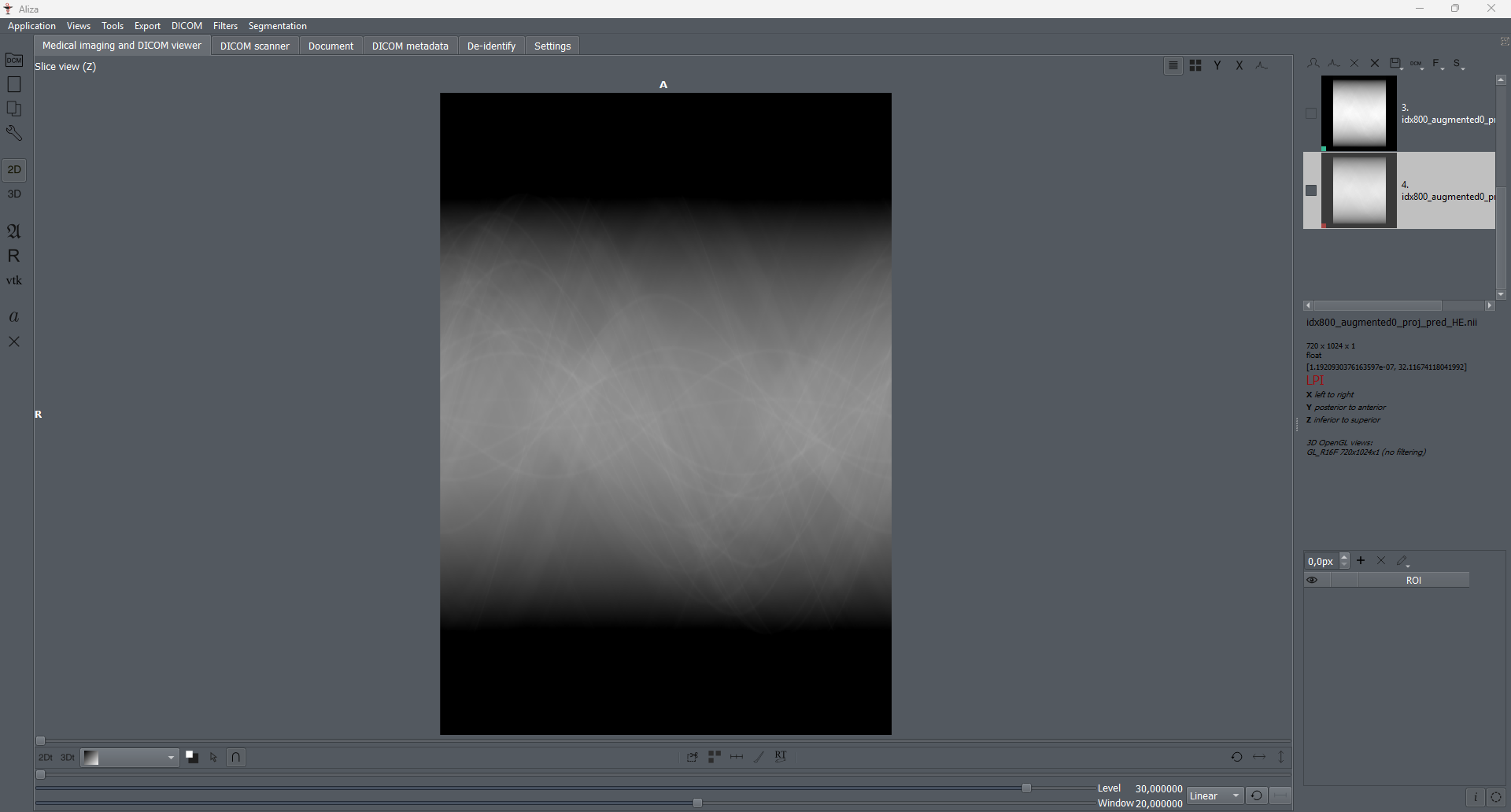}};
    \node[draw=none, left=of proj_HE, xshift=-1.2cm, font=\large, align=center, anchor=center] {Modelled \\ projection data $\hat{p}$};
    \draw[arrowgray] (Model.-80) |- (proj_HE.42) ;
    \draw[dashedarroworange] (proj_LE.-42) -| (Model.-100);

    % L2
    \node[ellipseboxorange, below=of Unet, yshift=-3.4cm] (L2) {\Large $\mathcal{L}$};
    \node[draw=none, right=of L2, xshift=-0.5cm, yshift=0.75cm, font=\huge] {\textcolor{orange}{$\nabla$}};
    
    % Dotted loss arrows
    \draw[dashedarrowgray] (measurements_HE) |- (L2);
    \draw[dashedarroworange, yshift=-1cm] (L2.10) -- (proj_LE.-137);
    \draw[dashedarrowgray] (proj_HE.137) -- (L2.-10);

    % \draw[fill=orange!20!, rectangle, below=of y, minimum height=0.5cm, yshift=-1cm, font=\huge] {\textcolor{orange!20!}{.}};
    
    % \draw[arrow, draw=orange!20!] (y) -- $(y) + (0, 5)$
    
    % Legende
    
    \draw[arrowgray] (0.0, -9.0) -- (0.7, -9.0);
    \node[draw=none, rectangle, align=left] at (2.0, -9.0) {\textcolor{gray}{Forward pass}};
    \draw[dashedarrowgray] (0.0, -9.5) -- (0.7, -9.5);
    \node[draw=none, rectangle, align=left] at (2.3, -9.5) {\textcolor{gray}{Loss calculation}};
    \draw[dashedarroworange] (0.0, -10.0) -- (0.7, -10.0);
    \node[draw=none, rectangle, align=left] at (2.1, -10.0) {\textcolor{orange}{Backward pass}};    
    
    % % black box as legend
    % \node[draw=gray, fill=none, rectangle, below=of y, minimum height=2cm, minimum width=10cm, xshift=2cm, yshift=-1cm, font=\huge] {\textcolor{mylightblue}{.}};
    
    \end{tikzpicture}
    } % end resizebox
    \caption{Training scheme visualized for single training data set \(p\) from the projection domain. Training is performed in batches of all training data.}
    \label{fig:AnsatzDECT}
\end{figure}
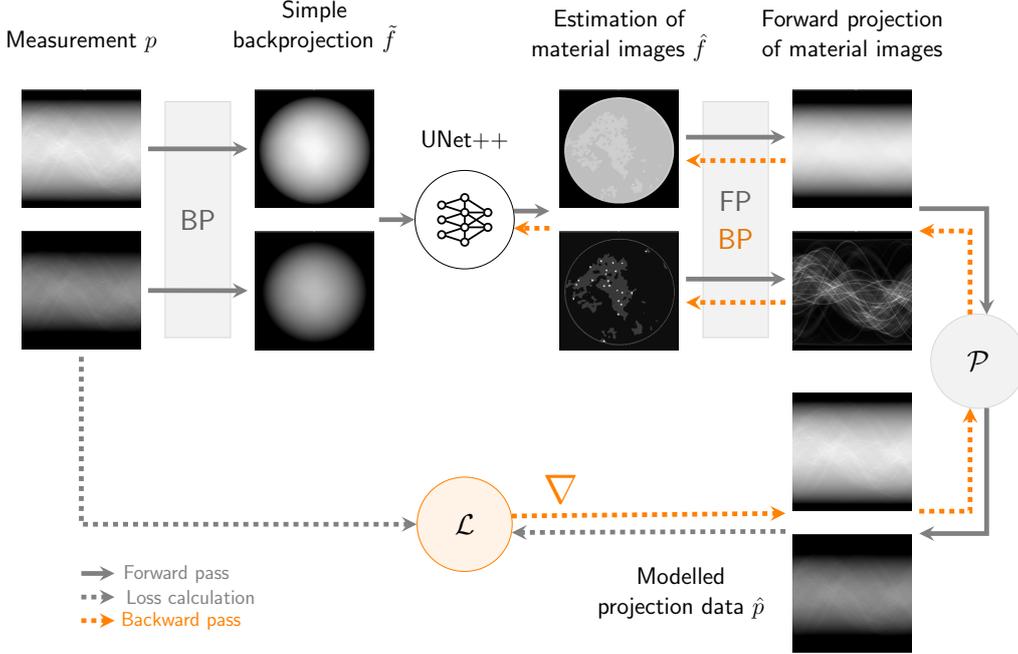

Our framework follows the general approach established in our previous work for CT reconstruction \cite{hellwege2024unsupervised}, adapted for the dual-energy material decomposition task. The pipeline consists of three stages (Figure~\ref{fig:AnsatzDECT}):

\begin{enumerate}
\item \textbf{Initial Estimate:} The dual-energy projection data $(p^{\text{LE}}, p^{\text{HE}})$ are transformed to the image domain via simple backprojection, providing initial estimates for the network input.

\item \textbf{Neural Network:} A deep convolutional neural network $\mathcal{N}_\theta$ with learnable parameters $\theta$ processes the initial estimates to produce material images $(\hat{f}^A, \hat{f}^B) = \mathcal{N}_\theta(\tilde{f}^{\text{LE}}, \tilde{f}^{\text{HE}})$.

\item \textbf{Polychromatic Forward Model:} The predicted material images are forward projected and passed through the polychromatic model $\mathcal{P}$ (equation~\eqref{eq:polychromatic}) to obtain predicted projections $(\hat{p}^{\text{LE}}, \hat{p}^{\text{HE}})$.
\end{enumerate}

\noindent As network $\mathcal{N}_\theta$ we employ the UNet++ architecture \cite{zhou2018unet++} with a ResNeXt-101 (32$\times$16d) encoder \cite{xie2017aggregated}, implemented using the Segmentation Models PyTorch library \cite{iakubovskii2019segmentation}. The network has two input and output channels corresponding to the two spectra and two basis material images respectively.

The network is trained to minimize the Mean-Squared-Error (MSE) between measured and predicted projections for both energy spectra:
\begin{equation}
\mathcal{L}(\theta) = \frac{1}{|\mathcal{D}|} \sum_{(p^{\text{LE}}, p^{\text{HE}}) \in \mathcal{D}} \frac{1}{M}  \sum_{i=1}^M \left( \| p^{\text{LE}}_i - \hat{p}^{\text{LE}}_i \|_2^2 + \| p^{\text{HE}}_i - \hat{p}^{\text{HE}}_i \|_2^2 \right),
\label{eq:loss}
\end{equation}
where $\mathcal{D}$ denotes a set of training data pairs. The loss function requires only the measured projection data and knowledge of the polychromatic forward model. Hence, no ground-truth material images are needed.
The gradient of the loss with respect to network parameters is computed via backpropagation through the entire pipeline, including the polychromatic forward model. The chain rule propagates gradients from the projection domain through the physics-based model to the network weights.

% \subsubsection{Training Procedure}

% Training was performed using the Adam optimizer \cite{kingma2014adam} with a learning rate of $10^{-4}$. We used a batch size of 8 samples, constrained by GPU memory (NVIDIA RTX A6000, 48\,GB VRAM). No data augmentation was applied to preserve physical consistency. Early stopping was employed based on validation loss, with training terminating when no improvement was observed for 100 consecutive epochs. The best-performing model on the validation set was saved for final evaluation.

Training was performed using the Adam optimizer \cite{kingma2014adam} with a base learning rate of $10^{-4}$ and a batch size of 8 samples, constrained by GPU memory (NVIDIA RTX A6000, 48\,GB VRAM). 
We employed a progressive training strategy to improve convergence stability: training began with a subset of 64 samples and was successively extended to 200, then 1000, and finally the full training set. At each stage, the network was initialized with weights from the previous stage. This incremental approach allows the network to first learn robust features from limited data before being exposed to the full variability of the dataset, which proved beneficial for the complex nonlinear mapping involving the polychromatic forward model.

A cyclic learning rate schedule was applied, starting each cycle at $10^{-3}$ and decaying to the base learning rate over 10 epochs. Early stopping was employed based on validation loss, with training terminating when no improvement was observed for 150 consecutive epochs. The best-performing model on the validation set was retained for final evaluation.

\subsection{Dataset}

We employed the AAPM DL-Spectral CT Challenge dataset \cite{aapm_spectral} for validation. This dataset was designed to advance DL methods for spectral CT imaging and includes simulated dual-energy projection data (low-energy and high-energy spectra) with corresponding ground-truth material images for evaluation of soft tissue, bone and calcifications. The materials soft tissue and calcification were chosen as basis materials for material decomposition.
The 1000 data sets were split into training (80\%, of which 20\% where used for validation) and test (20\%) sets, resulting in 200 test datasets for final evaluation. To increase data variability, the original phantom images were augmented through geometric transformations (five random combinations of rotations, flips and elastic deformations) applied in the image domain before prior to projection simulation. This ensured that the simulated projection data remains physically consistent with the augmented phantoms. Notably, augmentation was only applied to training data; the test set consists of original, non-augmented samples.

\subsection{Evaluation}
Since ground-truth material images are not available for direct comparison in clinical scenarios, we evaluate performance through two complementary approaches:
The predicted material images are forward projected through the polychromatic model, and the resulting projections are compared to the simulated raw data using root mean squared error (RMSE):
\begin{equation}
\text{RMSE} = \sqrt{\frac{1}{M} \sum_{i=1}^{M} (x_i - \hat{x}_i)^2},
\end{equation}
where $M$ is the total number of projection measurements across both energy spectra.
Virtual monoenergetic images (VMIs) are synthesized from the estimated decomposed material images $\hat{f}^A$ and $\hat{f}^B$ at the energies $E \in \{\text{30\,keV, 50\,keV, 70\,keV}\}$ using
\begin{equation}
\mu_{\text{VMI}}(E) = \mu^A(E) \hat{f}^A  + \mu^B(E) \hat{f}^B .
\end{equation}
These VMIs are compared against ground-truth VMIs using RMSE and structural similarity index (SSIM) \cite{wang2004image}. VMIs provide a clinically relevant evaluation metric as they represent the typical output used for diagnostic interpretation.

\section{Results}

\subsection{Projection Domain Results}

\begin{figure}
    \sffamily
    % \centering
    \hspace{1.5cm}
    \begin{tikzpicture}
        \node[anchor=south west] at (0,0){\includegraphics[width=0.7\linewidth, trim=1cm 0cm 0cm 0cm, clip]{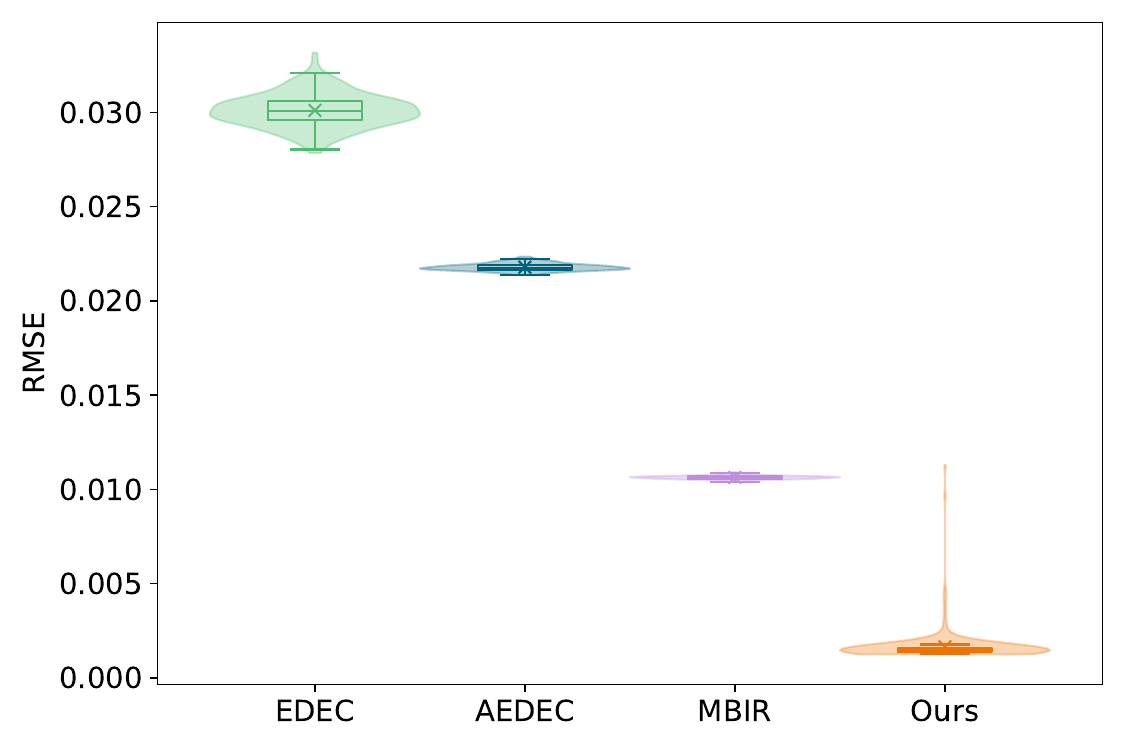}};
        \node[rotate=90] at (-0.2,3.5) {\footnotesize{RMSE}};
    \end{tikzpicture}
    \caption{Distributions of RMSE values in projection space for the 200 test datasets across different material decomposition methods.}
    \label{fig:AAPM_proj_quantitativ}
\end{figure}

We first analyze the discrepancies between the predicted and measured projection data for the LE and HE spectra. The predicted projections are computed from the material images of each method through the polychromatic forward model.
Figure~\ref{fig:AAPM_proj_quantitativ} shows the distribution of RMSE values across the 200 test datasets for each method. Clear differences between method distributions are evident. AEDEC, as an advancement of the basic EDEC approach, reduces discrepancies for all test datasets but is outperformed by the iterative MBIR approach. The unsupervised DL method shows few outliers in a similar value range as MBIR while further reducing the discrepancy to ground truth for most test datasets.

\begin{figure}
\sffamily
    \centering
    \begin{tikzpicture}
    
    \node[anchor=center] (img){\includegraphics[width=0.8\linewidth]{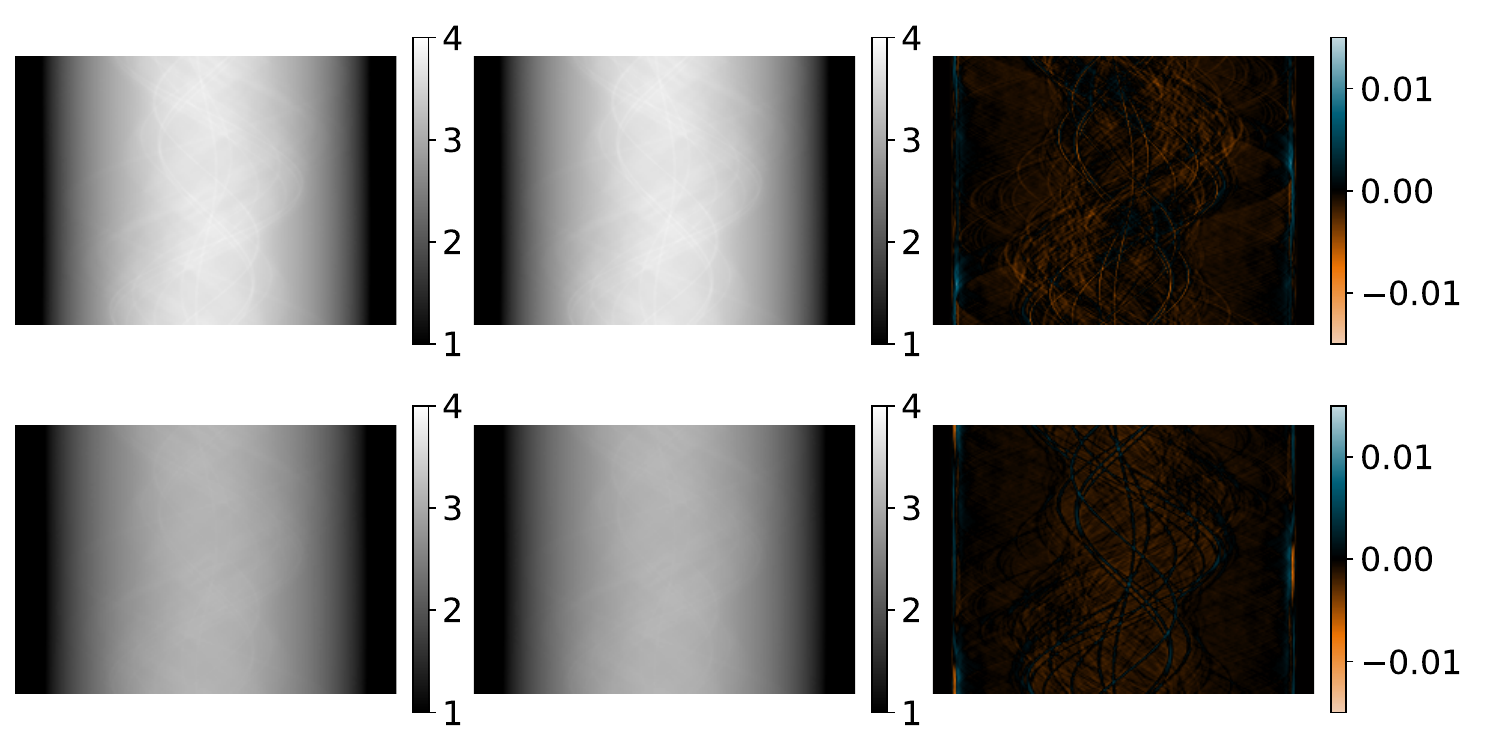}};

    \node[below=of img, yshift=+0.75cm] (img2){\includegraphics[width=0.8\linewidth]{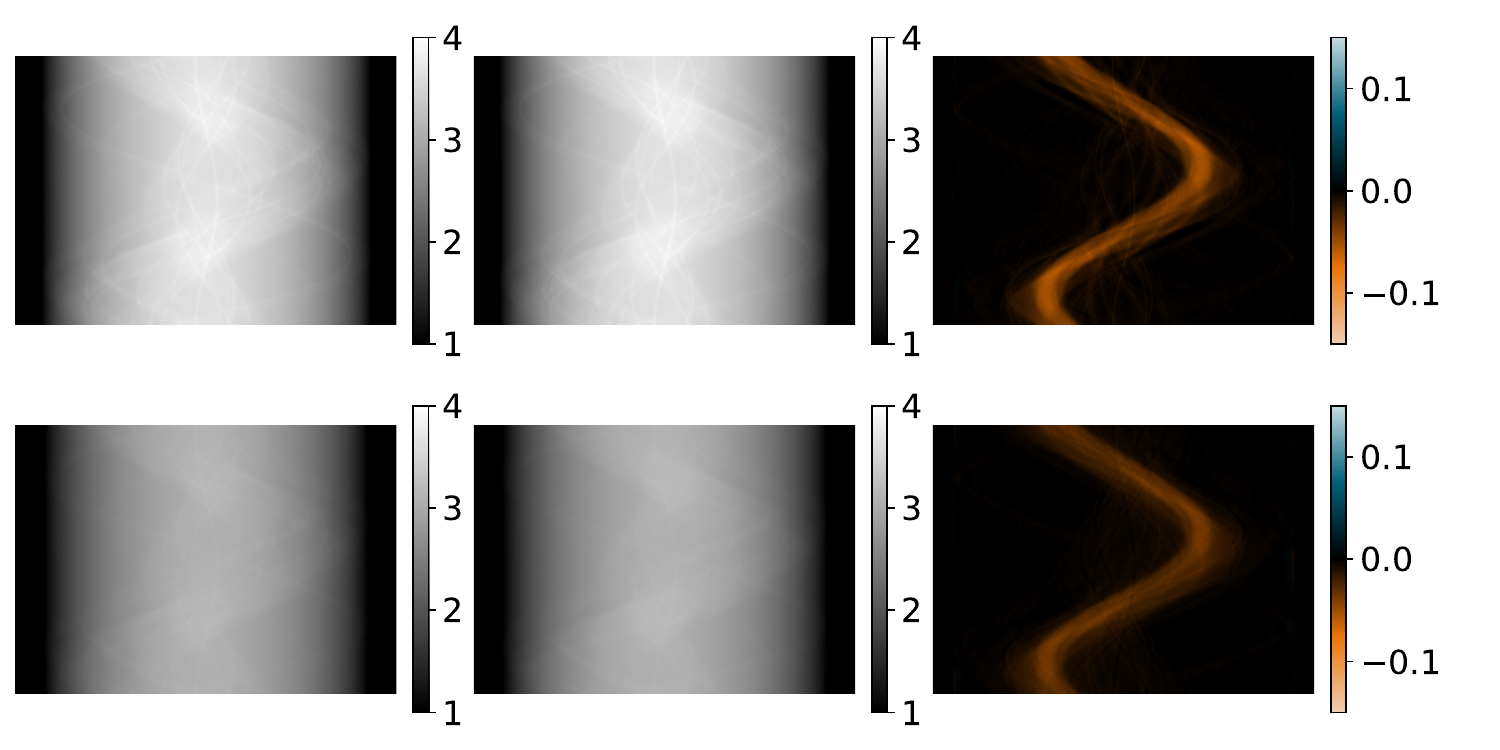}} ;

    \node at ([yshift=0cm, xshift=-1cm]img.north west) {(a)};
    \node at ([yshift=0cm, xshift=-1cm]img2.north west) {(b)};
    
    \node[rotate=90] at ([yshift=-1.8cm, xshift=-0.2cm]img.north west) {\small LE};
    \node[rotate=90]  at ([yshift=-4.5cm, xshift=-0.2cm]img.north west) {\small HE};

    \node[rotate=90]  at ([yshift=-1.8cm, xshift=-0.2cm]img2.north west) {\small LE};
    \node[rotate=90]  at ([yshift=-4.5cm, xshift=-0.2cm]img2.north west) {\small HE};

    \node at ([yshift=-0.2cm, xshift=+1.9cm]img.north west) {\small Prediction};
    \node at ([yshift=-0.2cm, xshift=-0.7cm]img.north) {\small GT};
    \node at ([yshift=-0.2cm, xshift=-3.2cm]img.north east) {\small Pred.$-$GT};
    
    \end{tikzpicture}
    
    \caption{Qualitative examples of computed LE/HE projection data from the network-predicted material images. (a) Example with low deviation from ground truth (RMSE\,=\,$1.29 \times 10^{-3}$). (b) Less accurate result (RMSE\,=\,$1.13 \times 10^{-2}$). All projection values are relative intensities (unitless).}
    \label{fig:AAPM_proj_qualitativ}
\end{figure}
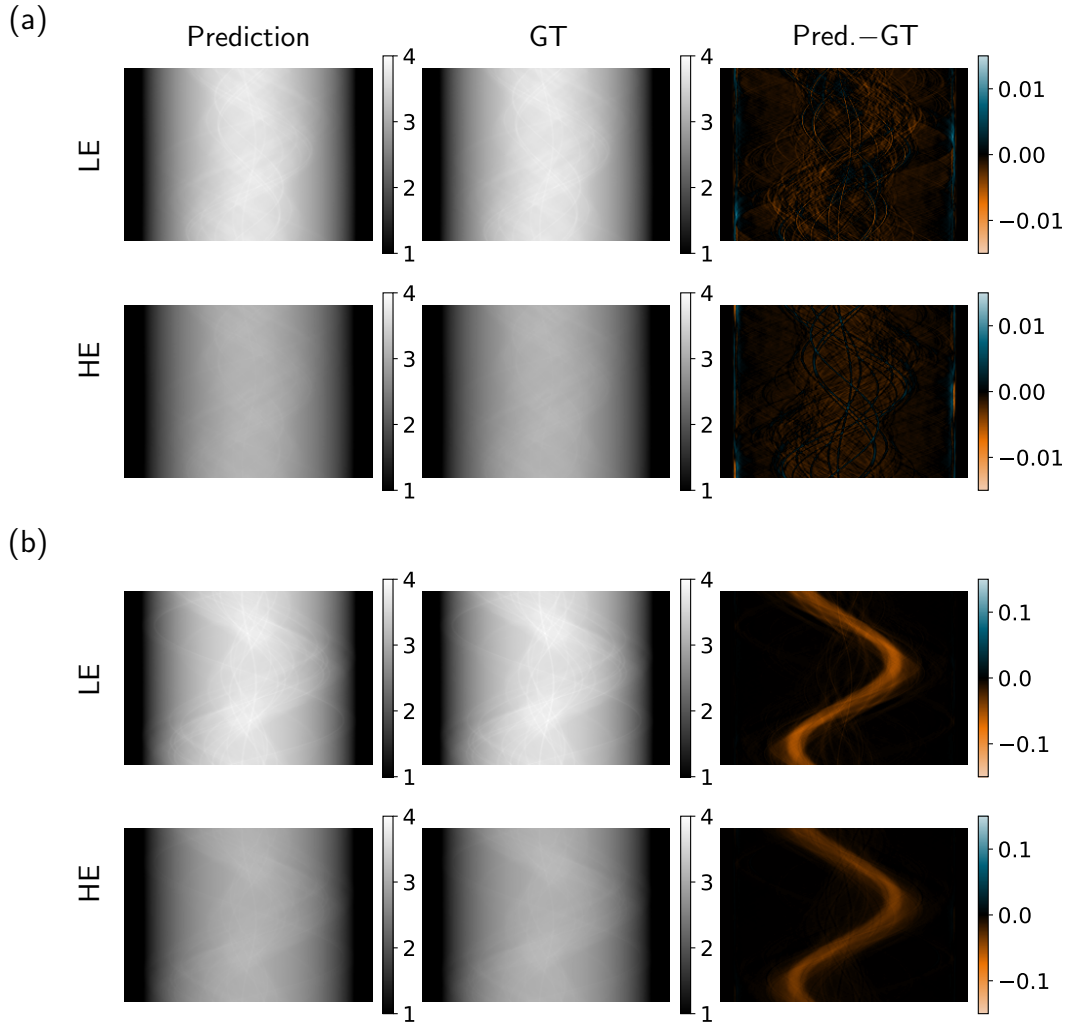

Qualitative examples are shown in Figure~\ref{fig:AAPM_proj_qualitativ} for two datasets representing good and poor agreement with the ground truth. The less accurately predicted sinogram in Figure~\ref{fig:AAPM_proj_qualitativ}(b) shows deviations primarily in regions of high absorption, i.e., ray paths through dense materials. The example of good agreement in Figure~\ref{fig:AAPM_proj_qualitativ}(a) exhibits only small deviations along edges in the sinogram. Overall, the deviations appear very small, which is confirmed by the quantitative RMSE values.

\subsection{Image Domain Results}

To evaluate material decomposition in the image domain, for which no direct ground truth exists, we analyze VMIs generated from the material images. Table~\ref{tab:results} shows the median RMSE and SSIM values for 200 estimated VMIs, each at three different energies (30\,keV, 50\,keV, 70\,keV), across the different methods.
All three conventional approaches are outperformed by the unsupervised DL method, which demonstrates clear improvements across all energies and both metrics in the form of lower medians for RMSE, higher medians for SSIM, and narrower interquartile ranges.
Table~\ref{tab:results} summarizes the median RMSE and SSIM values for all methods and energies. The unsupervised DL method achieves the lowest RMSE and highest SSIM across all evaluated VMI energies.
% \begin{figure}[ht]
%     \centering
%     \begin{tikzpicture}
        
%     \node[inner sep=0pt] (img30kV) {\includegraphics[width=0.8\linewidth]{eval_img_quantitative_30kV.pdf}};
        
%     \node[below=of img30kV, yshift=+1cm] (img50kV){\includegraphics[width=0.8\linewidth]{eval_img_quantitative_50kV.pdf}} ;
    
%     \node[below=of img50kV, yshift=+1cm] (img70kV){\includegraphics[width=0.8\linewidth]{eval_img_quantitative_70kV.pdf}};

%     \node at ([yshift=-0.6cm, xshift=+0.5cm]img30kV.north west) {(a) 30\,keV};
%     \node at ([yshift=-0.7cm, xshift=+0.5cm]img50kV.north west) {(b) 50\,keV};   
%     \node at ([yshift=-0.7cm, xshift=+0.5cm]img70kV.north west) {(c) 70\,keV};
    
%     \end{tikzpicture}
    
%     \caption{Distributions of RMSE and SSIM values (boxplots) for VMIs at (a) 30\,keV, (b) 50\,keV, and (c) 70\,keV for different material decomposition methods across the 200 test datasets.}
%     \label{fig:AAPM_img_quantitativ}
% \end{figure}

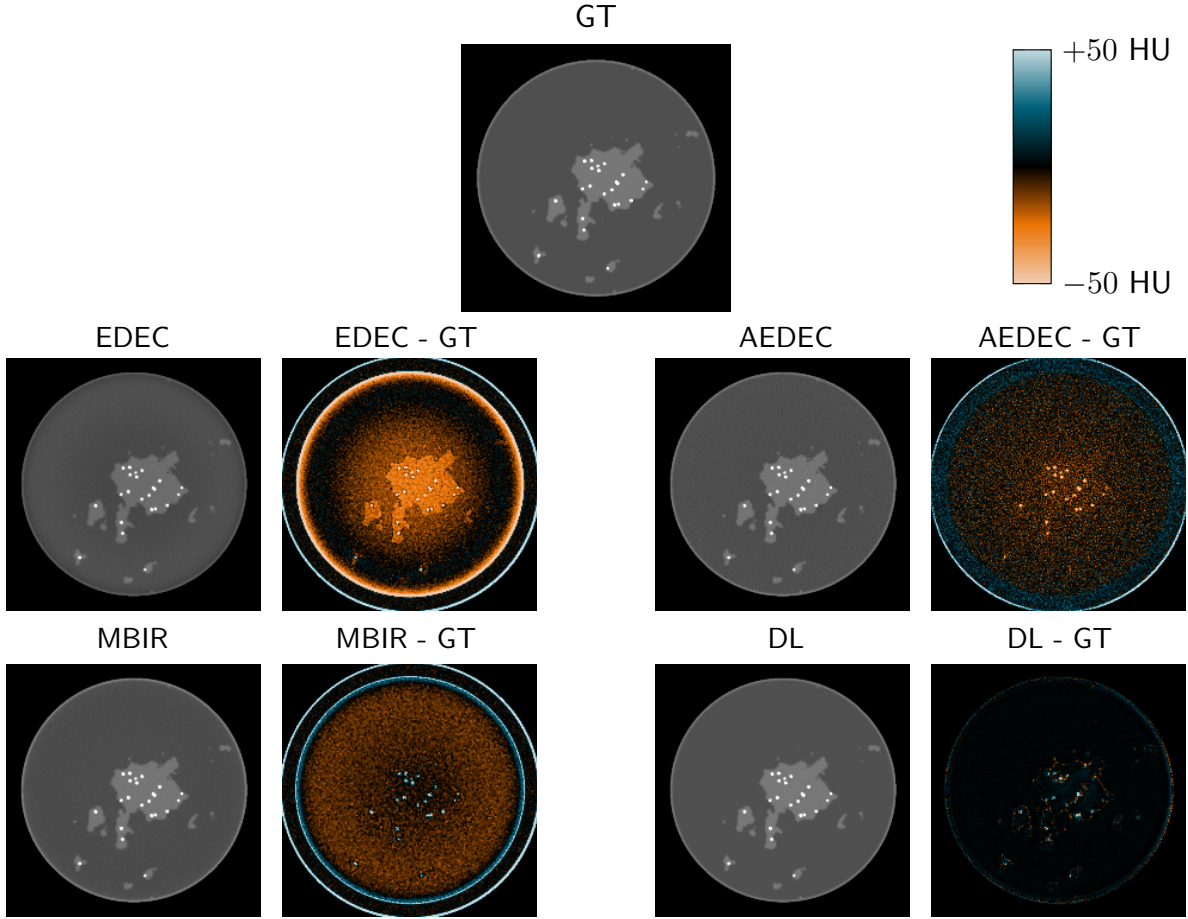
\begin{figure}[t]
\sffamily
    \centering
    \begin{tikzpicture}[scale=1.0,]
        \node[inner sep=0pt] (img) {\includegraphics[width=16cm, trim=1cm 0cm 1cm 0cm, clip]{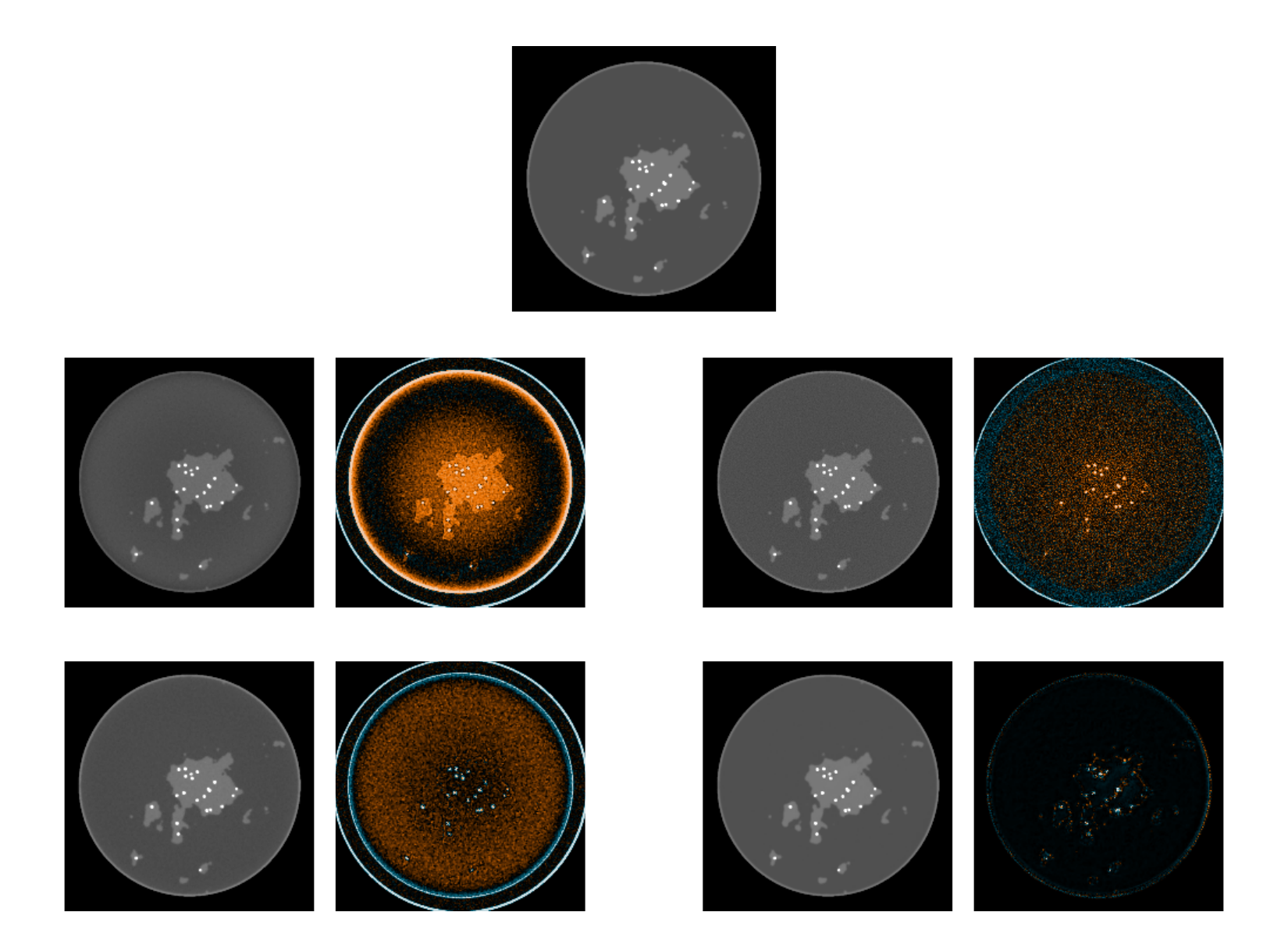}};
                
        \node at (img.north) [align=center, yshift=-0.25cm, xshift=0cm] {GT};
        
        % \node at ([yshift=-0.5cm, xshift=+2.5cm]img.north west) {\small VMI at $50\,\mathrm{keV}$};
        % \node at ([yshift=-0.5cm, xshift=+6.9cm]img.north west) {\small Diff.\ to GT};  
        % \node at ([yshift=-2.75cm, xshift=-0.5cm]img.north west) {\small EDEC};
        % \node at ([yshift=-7.25cm, xshift=-0.5cm]img.north west) {\small AEDEC};
        % \node at ([yshift=-11.0cm, xshift=-0.5cm]img.north west) {\small MBIR};
        % \node at ([yshift=-15.5cm, xshift=-0.5cm]img.north west) {\small DL};
        % \node at ([yshift=-19.5cm, xshift=-0.5cm]img.north west) {\small GT};

        \node at ([ yshift=-4.5cm, xshift=-6.1cm] img.north) {\small EDEC};
        \node at ([yshift=-4.5cm, xshift=-2.5cm] img.north) {\small EDEC - GT};
        \node at ([yshift=-4.5cm, xshift=+2.5cm] img.north) {\small AEDEC};
        \node at ([yshift=-4.5cm, xshift=+6.1cm]
        img.north) {\small AEDEC - GT};
        \node at ([yshift=-8.5cm, xshift=-6.1cm] img.north) {\small MBIR};
        \node at ([yshift=-8.5cm, xshift=-2.5cm] img.north) {\small MBIR - GT};
        \node at ([yshift=-8.5cm, xshift=+2.5cm] img.north) {\small DL};
        \node at ([yshift=-8.5cm, xshift=+6.1cm] img.north) {\small DL - GT};

        % Colorbar hinzufügen
        \node at ([yshift=-2.25cm, xshift=5.5cm] img.north){ % Positioniere die Colorbar hier
            \begin{axis}[
                anchor=north west,
                colorbar,
                colormap={custom_shift}{
                rgb255(0cm)=(240,205,179); 
                rgb255(1cm)=(236,116,4); 
                rgb255(2cm)=(0,0,0);
                rgb255(3cm)=(0,97,122); 
                rgb255(4cm)=(198,220,226)},
                colorbar style={
                    width=0.5cm,
                    height=3.1cm,
                    ytick={0,1,2,3,4},
                    yticklabels={
                    $-50 $ HU, 
                    $+50 $ HU},
                },
                axis y line=none, % Entferne die y-Achse
                axis x line=none, % Entferne die x-Achse
                xmin=0, xmax=1,
                ymin=0, ymax=1,
            ]
        \end{axis}
        };

    \end{tikzpicture}
     \caption{VMIs at 50\,keV for different material decomposition methods. The example dataset was selected as one with relatively high RMSE (RMSE\,=\,6.10\,HU) between the DL method result and ground truth (GT). VMIs are displayed in window [L: 0\,HU, W: 600\,HU].}
    \label{fig:AAPM_img_qualitativ}
\end{figure}

\begin{table}[t]
\caption{Mean RMSE (in HU) and SSIM values for VMIs at different energies. Best results are shown in bold.}
\centering
\begin{tabular}{l cc cc cc}
\toprule
 & \multicolumn{2}{c}{30\,keV} & \multicolumn{2}{c}{50\,keV} & \multicolumn{2}{c}{70\,keV} \\
\cmidrule(lr){2-3} \cmidrule(lr){4-5} \cmidrule(lr){6-7}
Method & RMSE & SSIM & RMSE & SSIM & RMSE & SSIM \\
\midrule
EDEC & 49.43 & 0.958 & 24.75 & 0.967 & 23.30 & 0.903 \\
& \hspace{0.3cm} $\pm$ 3.21 &\hspace{0.3cm} $\pm$ 0.007 &\hspace{0.3cm} $\pm$ 1.03 &\hspace{0.3cm} $\pm$ 0.005 &\hspace{0.3cm} $\pm$ 1.05 &\hspace{0.3cm} $\pm$ 0.015\\ 
AEDEC & 27.59  & 0.983  & 18.14  &0.985 & 18.51 & 0.966 \\
& \hspace{0.3cm} $\pm$ 1.73 & \hspace{0.3cm} $\pm$ 0.002 & \hspace{0.3cm} $\pm$ 0.31 & \hspace{0.3cm} $\pm$ 0.002  & \hspace{0.3cm} $\pm$ 0.26 & \hspace{0.3cm} $\pm$  0.004\\
MBIR & 31.24  & 0.977  & 13.07 & 0.988  & 13.32  & 0.981  \\
  & \hspace{0.3cm}  ± 1.64  & \hspace{0.3cm} ± 0.004  & \hspace{0.3cm}  ± 0.33  & \hspace{0.3cm}  ± 0.002  & \hspace{0.3cm} ± 0.28  & \hspace{0.3cm} ± 0.002 \\
\textbf{DL} & \textbf{12.64} & \textbf{0.997} & \textbf{5.83} & \textbf{0.999} & \textbf{5.88} & \textbf{0.997} \\
 & \hspace{0.3cm}± 0.70 & \hspace{0.3cm}  ± 0.001 & \hspace{0.3cm} ± 0.76 & \hspace{0.3cm}  ± 0.002 & \hspace{0.3cm} ± 0.43 & \hspace{0.3cm}  ± 0.003 \\

\bottomrule
\end{tabular}
\label{tab:results}
\end{table}
Similar trends are observed in the qualitative analysis shown in Figure~\ref{fig:AAPM_img_qualitativ}. The VMIs at the exemplary energy of 50\,keV show varying noise behavior in homogeneous regions. The HU values deviate for specific image regions or materials. For EDEC, bright regions in the image center and at object boundaries are noticeable. AEDEC shows reduced deviation in these regions but exhibits stronger noise in homogeneous areas. For MBIR, high deviations in the HU values of calcifications are particularly apparent. The VMI from the DL method also shows higher deviations in these regions of higher density; however, the image generally exhibits very good agreement with the ground truth.

\section{Discussion}

\subsection{Performance Analysis}

Our DL approach demonstrates excellent performance for DECT material decomposition on the AAPM Spectral CT dataset. The method significantly outperforms all conventional approaches (EDEC, AEDEC, MBIR) in both projection domain consistency and VMI accuracy across all evaluated energy levels.
The superior performance in the projection domain (Figure~\ref{fig:AAPM_proj_quantitativ}) indicates that the network has learned to produce material images that are highly consistent with the underlying physics. This is a direct consequence of our training objective, which minimizes projection-domain discrepancies. Importantly, this consistency translates to improved image-domain performance as measured through VMI evaluation.

% The observation that AEDEC performs comparably to MBIR at low energies (30\,keV) is noteworthy. Low-energy VMIs amplify differences in material composition but are also more sensitive to noise and decomposition errors. The empirical calibration approach of AEDEC appears to provide robust performance in this regime, while MBIR without regularization may be affected by noise amplification.
% The consistent improvement of our DL method across all energies suggests that the network learns a robust decomposition mapping that generalizes well. The narrower distributions in Figure~\ref{fig:AAPM_img_quantitativ} indicate reduced variance across different anatomical configurations and material compositions.\\

% \subsection{Advantages of unsupervised Training}
% The primary advantage of our approach is the elimination of ground-truth material images during training. 
Compared to existing DL approaches for DECT material decomposition \cite{clark2018deep, xu2021image}, our method does not require supervised training. 
% This removes a significant practical barrier to clinical deployment, as ground-truth data acquisition is eliminated.
% In clinical DECT, obtaining true material ground truth is practically impossible since material concentrations cannot be directly measured, and phantom-based approaches do not represent the full variability of patient anatomy and pathology. 
In clinical DECT, obtaining true material ground truth is not possible in general, since the decomposition coefficients $f^A$ and $f^B$ are abstract quantities that depend on the choice of basis materials and have no direct physical measurement equivalent. Only derived quantities such 
as VMIs, which map back to energy-dependent attenuation coefficients, can be validated against physical reference data (e.g., NIST tables). Our approach circumvents this fundamental limitation by operating entirely in the projection domain, where measured data provides the ground truth.
With our method, we pave the way for training on clinical datasets without the need for paired ground-truth images or extensive phantom calibrations. The network learns the inverse mapping purely from consistency with the physics-based forward model.
% \subsection{Comparison with Related Work}
% Our approach extends the concept of physics-informed unsupervised learning, which we previously demonstrated for single-energy CT reconstruction \cite{hellwege2024unsupervised}, to the more challenging problem of dual-energy material decomposition. The key extension is the incorporation of the polychromatic forward model, which introduces nonlinearity in the relationship between material images and projections.
Compared to MBIR, our approach offers substantially reduced inference time once trained. While MBIR requires iterative optimization for each dataset, our network produces material images in a single forward pass after completed training. The computational advantage becomes more pronounced for large-scale clinical applications or real-time imaging scenarios.\\
\vspace{11pt}

\subsection{Limitations}
The AAPM dataset uses simulated data with known ground truth for evaluation. While this enables rigorous quantitative assessment, validation on clinical DECT data will strengthen the clinical relevance of our findings. It should be noted that the simulation of measured data and reconstruction have been performed with the same polychromatic model, which poses an significant advantage over typical settings, where the model has to be estimated.
The current implementation assumes fixed projection geometry. Extension to variable geometries or different scanner configurations would require retraining or geometry-adaptive network architectures. This work focuses on 2D slice-by-slice processing. Extension to fully 3D volumetric decomposition will better exploit spatial correlations but requires substantially more computational resources.

% \textbf{Noise Characteristics:} As observed in the qualitative results, the DL method may introduce subtle smoothing effects. Investigation of noise texture and its impact on diagnostic tasks warrants further study.
% Future work will address these limitations by:
% \begin{itemize}
% \item Validating on clinical DECT acquisitions
% \item Developing geometry-adaptive network architectures
% \item Extending to 3D cone-beam and photon-counting CT
% \item Incorporating uncertainty quantification for clinical decision support
% \end{itemize}

\section{Conclusion}

We have presented a DL framework for dual-energy CT material decomposition that eliminates the requirement for ground-truth material images during training. By incorporating a polychromatic forward model into the training pipeline, the network learns to produce material images that are consistent with measured projection data.
Validation on the AAPM DL-Spectral CT Challenge dataset demonstrates that our method significantly outperforms conventional approaches including EDEC, AEDEC, and MBIR. The unsupervised DL method achieves the lowest RMSE and highest SSIM for virtual monoenergetic images across all evaluated energies (30\,keV, 50\,keV, 70\,keV).
These results establish unsupervised DL as a promising approach for DECT material decomposition, with significant practical advantages for clinical deployment where ground-truth training data is unavailable. The framework opens pathways toward accurate, efficient material decomposition that can be trained directly on clinical acquisitions.

\subsubsection*{Acknowledgements}
The authors thank the AAPM for providing the DL-Spectral CT Challenge dataset. Generative AI tools were used to assist with manuscript preparation. All content was reviewed and verified by the authors.

\subsubsection*{Funding}
This work was partially funded by the state of Schleswig-Holstein through the project ``Individualisierte Medizintechnik für bildgestützte, robotische Interventionen (IMTE 2)'', project number: 125 24 009.

% \roles{L.H.: Conceptualization, Methodology, Software, Validation, Original Draft. J.C.E.: Software, Visualization,  Review \& Editing. M.Sch.: Validation, Review \& Editing. T.M.B.: Supervision, Review \& Editing. M.St.: Conceptualization, Supervision, Review \& Editing.}
\subsubsection*{Contributions}
L.H. and M.St. conceived and designed the study. L.H. developed the methodology, implemented the software, and performed the validation experiments. J.C.E. contributed to software development and helped creating the visualizations. M.Sch. contributed to validation experiments. T.M.B. and M.St. supervised the research. L.H. wrote the original draft. All authors reviewed and edited the manuscript and approved the final version.

\subsubsection*{Data}
The AAPM DL-Spectral CT Challenge dataset is publicly available through the AAPM.

\end{document}